\documentstyle[12pt,a4wide]{article}
\input{epsf}
\newcommand{\be}{\begin{equation}}
\newcommand{\ee}{\end{equation}}
\newcommand{\bea}{\begin{eqnarray}}
\newcommand{\eea}{\end{eqnarray}}
\newcommand{\bean}{\begin{eqnarray*}}

\newcommand{\eean}{\end{eqnarray*}}
\font\upright=cmu10 scaled\magstep1 \font\sans=cmss10
\newcommand{\ssf}{\sans}
\newcommand{\stroke}{\vrule height8pt width0.4pt depth-0.1pt}
\newcommand{\Z}{\hbox{\upright\rlap{\ssf Z}\kern 2.7pt {\ssf Z}}}

\newcommand{\C}{{\rlap{\rlap{C}\kern 3.8pt\stroke}\phantom{C}}}
\newcommand{\R}{\hbox{\upright\rlap{I}\kern 1.7pt R}}
\newcommand{\CP}{\C{\upright\rlap{I}\kern 1.5pt P}}
\newcommand{\PP}{\hbox{\upright\rlap{I}\kern 1.5pt P}}

\newcommand{\identity}{{\upright\rlap{1}\kern 2.0pt 1}}

\newcommand{\pp}{\Delta}
\newcommand{\HH}{\mbox{\hbox{\upright\rlap{I}\kern 1.7pt H}}}

\newcommand{\zb}{{\bar z}}

\newcommand{\fr}{\frac}
\newcommand{\lm}{\lambda}
\newcommand{\ra}{\rightarrow}
\newcommand{\al}{\alpha}
\newcommand{\bt}{\beta}
\newcommand{\pr}{\partial}
\newcommand{\hs}{\hspace{5mm}}

\newcommand{\dg}{\dagger}

\newcommand{\acc}{\\[3mm]}

\begin{document}
\begin{center}{\bf
$SU(5)$ Gravitating  Monopoles }\\
\vglue 0.5cm { Yves Brihaye$^\dg${\footnote{{\it Email}:
yves.brihaye@umh.ac.be}}
and Theodora Ioannidou$^\ddagger${\footnote{{\it Email}:
ti3@auth.gr}}
} \\
\vglue 0.3cm
$^\dg${\it
Facult\'e des Sciences, Universit\'e de Mons, 7000 Mons, Belgium\\
$^\ddagger${\it Maths Division, School of Technology, University of
Thessaloniki, Thessaloniki 54124, Greece
}\\
}
\end{center}

\begin{abstract}
Spherically symmetric solutions of the $SU(5)$
Einstein-Yang-Mills-Higgs system
are constructed  using the harmonic map ansatz \cite{IS}.
This way the problem  reduces  to solving  a set of ordinary
differential equations  for the appropriate profile functions.
\end{abstract}

\section{Introduction}
Magnetic monopoles are of diverse interest since they are
predicted from grant unified theories (GUT) and embody a rich
mathematical structure. Also, they appear in non-perturbative field
theories and  provide a new perspective on particle physics
phenomenology. In particular,  the  $SU(5)$ gauge group plays a
central role in GUT
 and thus
it is natural to classify the magnetic monopoles related to this
model \cite{dt1}. 
A few years ago, the effects of gravitation on
monopoles were (also) considered \cite{bfm} and 
revealed a rich pattern of solutions (including the occurrence of
black holes) 
related  to the gravitational
parameter: $\alpha^2 \equiv G v$ where $G$ denotes Newton constant
 and  $v\in \R$ is  the vacuum expectation value of the Higgs field.
More recently, a new interest for $SU(5)$ monopoles
was stimulated by the discovery of a deep analogy between
their magnetic charges and  the electric charges in one
generation of elementary particles \cite{Vach}. This originated
several new papers on the topic, see e.g. \cite{lepora,pogosian}
and references therein. Here we use the
harmonic map ansatz \cite{IS}, recently applied to  $SU(3)$
gravitating monopole \cite{BHIZ}, in order to construct their
$SU(5)$ counterparts. 
A similar
analysis has been applied  for deriving  $SU(5)$ solutions
(including black holes) which are {\it embeddings} of the $SU(2)$
ones \cite{BH} . However, the  solutions constructed here are
{\it non-embedded} of the $SU(2)$ ones and correspond to
monopole-antimonopole configurations.

The $SU(5)$ Einstein-Yang-Mills-Higgs action is given by:
 \be
  S=\int \left[\fr{R}{16\pi G}-\fr{1}{2} \mbox{tr}\left(F_{\mu
  \nu}\,F^{\mu  \nu}\right)-\mbox{tr}\left(D_\mu \Phi\, D^\mu
  \Phi\right)- V(\Phi)
\right]\sqrt{-g}\, d^4x
  \label{ac}
  \ee
where the potential is of the form \cite{Vach}:
\be
V(\Phi)=-\lm_1 \mbox{tr}(\Phi^2)+\lm_2\left(\mbox{tr}(\Phi^2)\right)^2
+\lm_3
\mbox{tr}(\Phi^4)-V_{min}.
\ee
Here   $g$ denotes the determinant of the
metric while the field strength tensor is defined by:
$F_{\mu
  \nu}=\pr_\mu A_\nu-\pr_\nu A_\mu+[A_\mu,A_\nu]$
and the
  covariant derivative of the Higgs field reads: $
D_\mu \Phi=\pr_\mu
  \Phi+[A_\mu,\Phi].$
The matrix $\eta$ represents a constant matrix of the form:
$\eta=i v {\bf 1}_N$, where ${\bf 1}_N$ denotes the unit matrix
in $N$ dimensions. Finally,  
$V_{min}=-15\lm_1^2/(60\lm_2+14\lm_3)$ has been subtracted due to
the finiteness of the energy.

The boundary conditions are such that the energy is finite and the
Higgs field at infinity is a given constant matrix
$\Phi(0,0,\infty)=i\Phi_0$ \be \Phi_0=\mbox{diag}
\left(\kappa_1,\kappa_2,\kappa_3,\kappa_4,\kappa_5\right) \ee in a
chosen direction (while since $\Phi \in su(5)$: $\sum_{i}^5
\kappa_i=0$). 
In addition, the asymptotic values $G_0$ of  the magnetic
charge $G=(1+|z|^2)^2F_{z\bar{z}}$ is given by 
\be 
G_0=G(0,0,1)=\mbox{diag}
\left(n_1,n_2-n_1,n_3-n_2, n_4-n_3,-n_4\right).
 \ee

Variation of  (\ref{ac}) with respect to the metric
$g^{\mu \nu}$ leads to the Einstein equations
 \be
 R_{\mu \nu}-\fr{1}{2}g_{\mu \nu} R=8\pi G \,T_{\mu \nu}
 \label{En}
 \ee
 with the stress-energy tensor
 $T_{\mu\nu}=g_{\mu\nu}{\cal L}-2\fr{\pr{\cal L}}
{\pr g^{\mu\nu}}$ given by
 \bea
 T_{\mu \nu}\!\!\!&=&\!\!\!\mbox{tr}\left(2D_\mu \Phi\, D_\nu \Phi-
 g_{\mu \nu} D_\al\Phi\,D^\al \Phi\right)\!
 +\!2\mbox{tr}\left(g^{\al \bt}\,F_{\mu \al}F_{\nu \bt}-\fr{1}{4}
g_{\mu \nu}\,F_{\al \bt}F^{\al \bt}\right)-g_{\mu\nu}V(\Phi).\hs\hs
 \eea
 In what follows we consider the static Einstein-Yang-Mills-Higgs equations
in order to construct  their $SU(5)$ spherically symmetric and purely magnetic
 (ie $A_0=0$)  solutions based on the harmonic map ansatz first introduced in
\cite{IS}.

\section{Spherical Symmetry}

The starting point of our investigation is the introduction of the
coordinates $r,z,\zb$ on $\R^3$. In terms of the usual spherical
coordinates $r,\theta,\phi$ the Riemann sphere variable $z$ is
given by $z=e^{i\phi} \tan(\theta/2)$. 
 In this system of coordinates the
Schwarzschild-like metric reads:
\begin{equation}
\label{metric}
ds^2=-A^2(r)B(r)dt^2+\fr{1}{B(r)}dr^2+\frac{4r^2}{(1+|z|^2)^2}
\,dz d\bar{z}, \hs B(r)=1-\frac{2m(r)}{r},
 \label{s}
\end{equation}
where  $A$ and $B$ are the metric  functions which are real
and depend only on the radial coordinate $r$, and
$m(r)$ is the mass function.
 The (dimensionfull) mass of the solution
is given by $m_{\infty}\equiv m(\infty)$.
For this metric the square-root of the determinant takes the simple form:
\begin{equation}
\sqrt{-g}=iA(r)\,\frac{2r^2}{(1+|z|^2)^2}.
\end{equation}
Then, the action  (\ref{ac}) simplifies to
\bea
S\!\!\!&=&\!\!\!\!\!\int
\bigg\{ \fr{B(1+|z|^2)^2}{r^2}\mbox{tr}\left(|F_{rz}|^2\right)+
\fr{(1+|z|^2)^4}{4r^4}\mbox{tr}\left(F_{z\bar{z}}^2\right) -
B \,\mbox{tr}\left( (D_r \Phi)^2\right)
-\fr{(1+|z|^2)^2}{r^2}\mbox{tr}\left(|D_z\Phi|^2\right)\nonumber \\
&&\,\,\,\,\,\,-V(\Phi)\bigg\} \sqrt{-g}\,r^2\,drdt
\label{ac1}
\eea
and the matter equations can be obtained by its variation
with respect to the matter fields.

 In addition, the Einstein equations (\ref{En}) take the form:
 \begin{equation}
 \frac{2}{r^2}\,m' = 8\pi
 G\, T^0_0,\hs  \hs
 \fr{2}{r} \fr{A'}{A}\,B =
8\pi G \,\left( T^0_0-T^r_r \right)
\label{ff}
 \end{equation}
where  prime denotes the derivative with respect to $r$, and
\bea
&&\!\!\!\!\!\!\!\!\!T_0^0=\fr{(1+|z|^2)^4}{4r^4}\mbox{tr}
\left(F_{z\bar{z}}^2\right)
-\fr{B(1+|z|^2)^2}{r^2}\mbox{tr}
\left(|F_{rz}|^2\right) -
B\, \mbox{tr}\left( (D_r \Phi)^2\right)
-\fr{(1+|z|^2)^2}{r^2}\mbox{tr}\left(|D_z\Phi|^2\right)
-V(\Phi)\hs\hs\hs\hs\nonumber\acc
&&\!\!\!\!\!\!\!\!\!T^0_0-T^r_r=
-\frac{2B(1+|z|^2)^2}{r^2}\mbox{tr}\left(|F_{rz}|^2\right)
- 2B\, \mbox{tr}\left((D_r\Phi)^2\right).\hs\hs
 \eea

Next we introduce the harmonic map ansatz for
 the Higgs and gauge fields  \cite{IS}
 \be
 \Phi=i\sum_{j=0}^{3} h_j\left(P_j-\fr{1}{N}\right),\hs \hs
 A_z=\sum_{j=0}^{3}g_j\left[P_j,\pr_z P_j\right],\hs
 A_r=0 \label{f}
 \ee
 where $h_j(r)$, $g_j(r)$ are the radial depended matter profile functions
and $P(z,\zb)$
 are $N\times N$ Hermitian  projectors: $P_j=P_j^\dg=P_j^2$,
 which are independent of the radius $r$.
 Note that all $N-1$ projectors $P_i$ are  orthogonal to each other since
 $P_iP_j=0$ for $i \neq j$ and
that we are working in a real gauge, since
 $A_{\zb}=-A_z^\dg$.
As shown in \cite{IS}, the projectors $P_k$ defined as
\be P_k=\fr{(\pp^k f)^\dg \pp^k f}{|\pp^k f|^2},
 \hs \hs k=0,..,N-1
\ee
where $  \pp f=\pr_z f- \fr{f \,(f^\dg\,\pr_z f)}{|f|^2}$
give the required set of orthogonal harmonic maps (for
details see \cite{Za}).
Moreover, the  spherically symmetric harmonic maps 
can be constructed by applying the orthogonalization procedure to the
initial holomorphic vector 
\be f=(1,2z,\sqrt{6}z^2,2z^3,z^4)^\dg.
\label{smap}
\ee
Then, under the transformation:  $
  h_j=\sum_{k=j}^{3}b_k$ and $c_j=1-g_j-g_{j+1}$ for $j=0,\dots,3$
and $g_{4}=0$,  the  equations of the  profile functions
$b_0,b_1,b_2,b_3$ and $c_0,c_1,c_2,c_3$ can be obtained from
variation of (\ref{ac1}). In fact, the energy-momentum tensor
$T_0^0$ can be evaluated explicitly:
\bea
T_0^0\!\!\!&=&\!\!\!\fr{4B}{r^2}\left[c_0'^2+\fr{3}{2}\left(c_1'^2+c_2'^2\right)
+c_3'^2\right]
+\fr{4}{r^2}\left[c_0^2b_0^2+\fr{3}{2}\left(c_1^2b_1^2+c_2^2b_2^2\right)+
c_3^2b_3^2\right]\nonumber\\
&+&\!\!\!\fr{4B}{5}\left[b_0'^2+\fr{3}{2}\left(b_1'^2+b_2'^2\right)+b_3'^2
+b_0'\left(
b_2'+\fr{3b_1'}{2}+\fr{b_3'}{2}\right)+b_3'\left(b_1'+\fr{3}{2}b_2'\right)
+2b_1'b_2'\right]\nonumber\\
&+&\!\!\!\fr{1}{r^4}\left[8c_0^4+18\left(c_1^4+c_2^4\right)+8c_3^4-4c_0^2-6
\left(c_1^2+c_2^2\right)-
4c_3^2-12\left(c_1^2c_0^2+c_2^2c_3^2\right)-18c_1^2c_2^2+10\right]\nonumber\\
&-&\!\!\!V(\Phi)
\label{ene}
\eea
where
\bea
V(\Phi)\!\!\!\!&=&\!\!\!\!\fr{4\lm_1}{5}\left[b_0^2+\fr{3}{2}
\left(b_1^2+b_2^2\right)+b_3^2+
b_0\left(b_2+\fr{3b_1}{2}+\fr{b_3}{2}\right)+b_3\left(b_1+\fr{3b_2}{2}\right)+
2b_1b_2\right]\nonumber\\
&+&\!\!\!\!\fr{16\lm_2}{25}\left[b_0^2+\fr{3}{2}\left(b_1^2+b_2^2\right)+b_3^2+
b_0\left(b_2+\fr{3b_1}{2}+\fr{b_3}{2}\right)+b_3\left(b_1+\fr{3b_2}{2}\right)+
2b_1b_2\right]^2\nonumber\\
&+&\!\!\!\!\fr{\lm_3}{125}\Bigg\{52\left[b_0^3\left(b_3\!+\!3b_1
\!+\!2b_2\right)\!+\!b_3^3\left(b_0\!+\!3b_2\!+\!2b_1\right)\right]
\!+\!56\left[b_1^3\left(b_3\!+\!\fr{3b_0}{2}\!+\!2b_2\right)\!+\!
b_2^3\left(b_0\!+\!\fr{3b_3}{2}\!+\!2b_1\right)\right]\nonumber\\
&&\hs +42\left(b_2^4+b_1^4\right)\!+\!52
\left(b_0^4+b_3^4\right)\!+\!12\left[b_0b_1\left(11b_0+7b_1\right)
\left(b_3+2b_2\right)
\!+\!b_2b_3\left(11b_3+7b_2\right)\left(b_0+2b_1\right)\right]\nonumber\\
&&\hs +192b_2b_1\left(b_2+b_3\right)\left(b_0+b_1\right)+108\left[b_0^2b_2\left(
b_2+b_3\right)+b_3^2b_1\left(b_1+b_0\right)\right]
+198\left(b_2^2b_3^2+b_0^2b_1^2\right)\nonumber\\
&&\hs +42b_0^2b_3^2\Bigg\}-V_{min}.
\label{pot}
\eea
It can be seen  that the energy is finite providing
the functions approach their asymptotic
values at least as fast as $1/r$, and if (in addition) the constraints:
$c_j(\infty)b_j(\infty)=0$ are imposed, for all $j$.

In order to read off the properties of a given solution we need to compute
the Higgs field and magnetic charge  at $z=0$. Explicitly, these are
 given by
\bea
\Phi_0\!\!\!\!&=&\!\!\!\!\fr{1}{5}\,\mbox{diag}\bigg(\!
b_3\!+\!3b_1\!+\!2b_2\!+\!4b_0,
b_3\!+\!3b_1\!+\!2b_2\!-\!b_0,
b_3\!-\!b_0\!-\!2b_1\!+\!2b_2,\nonumber\\
&&\hs\hs \hs b_3\!-\!2b_1\!-\!3b_2\!-\!b_0,
-4b_3\!-\!2b_1\!-\!3b_2\!-\!b_0\bigg)\acc\nonumber\\
G_0\!\!\!\!&=&\!\!\!\!\mbox{diag}\left(4\left(1-c_0^2\right),
 2\left(1+2c_0^2-3c_1^2\right),
6\left(c_1^2-c_2^2\right),-2(1+2c_3^2-3c_2^2),
-4\left(1-c_3^2\right)
\right)\hs
\eea
 which (also) determine the boundary conditions of the matter 
 profile functions.

After some algebra, it can be shown that the Higgs profile functions satisfy
the following ordinary differential equations:
 \bea
\fr{1}{A}\left(AB\,c_0'\right)'&=&
b_0^2\,c_0+\fr{1}{r^2}\,c_0\left(4c_0^2-3c_1^2-1\right),
\nonumber\acc
\fr{1}{A}\left(AB\,c_1'\right)'&=&
b_1^2\,c_1+\fr{1}{r^2}\,c_1\left(6c_1^2-2c_0^2-3c_2^2-1\right),
\nonumber\acc
\fr{1}{A}\left(AB\,c_2'\right)'&=&
b_2^2\,c_2+\fr{1}{r^2}\,c_2\left(6c_2^2-2c_3^2-3c_1^2-1\right),
\nonumber\acc
\fr{1}{A}\left(AB\,c_3'\right)'&=&
b_3^2\,c_3+\fr{1}{r^2}\,c_3\left(4c_3^2-3c_2^2-1\right)
 \eea
while the  profile functions of the  the gauge fields satisfy:
\bea
\fr{\left(r^2\,AB\,b_0'\right)'}{2Ar^2}
\!\!\!\!&=&\!\!\! \fr{1}{r^2}\left(4b_0c_0^2-3b_1c_1^2\right)-\fr{1}{2}\lm_1b_0
\hs\nonumber\\
&-&\!\!\!
\fr{4\lm_2}{5}\,b_0\left[b_0^2+\fr{3}{2}\left(b_1^2+b_2^2\right)+b_3^2+
b_0\left(b_2+\fr{3b_1}{2}+\fr{b_3}{2}\right)+b_3\left(b_1+\fr{3b_2}{2}\right)+
2b_1b_2\right]\nonumber\\
&-&\!\!\!\fr{\lm_3}{25}\,b_0\left[13b_0^2+27b_1^2+12b_2^2+3b_3^2+18b_0
\left(b_2-\fr{3b_1}{2}+\fr{b_3}{2}\right)+18b_1\left(b_3+2b_2\right)+12b_2 b_3
\right],\nonumber\acc
\fr{\left(r^2\,AB\,b_1'\right)'}{2Ar^2}
\! \!\!\!&=&\! \!\!\! \fr{1}{r^2}\left(6b_1c_1^2-3b_2c_2^2-2b_0c_0^2\right)-
\fr{1}{2}\lm_1b_1\nonumber\\
&-&\!\!\!
\fr{4\lm_2}{5}\,b_1\left[b_0^2+\fr{3}{2}\left(b_1^2+b_2^2\right)+b_3^2+
b_3\left(b_1+\fr{3b_2}{2}+\fr{b_0}{2}\right)+b_0\left(b_2+\fr{3b_1}{2}\right)+
2b_1b_2\right]\nonumber\\
&-&\!\!\!\fr{\lm_3}{25}\,b_1\left[3b_3^2+12b_2^2+7b_1^2+3b_0^2+6b_3
\left(2b_2+\fr{b_1}{2}-b_0\right)-3b_0\left(4b_2+b_1\right)+6b_1 b_2
\right],\nonumber\acc
\fr{\left(r^2\,AB\,b_2'\right)'}{2Ar^2}
\! \!\!\!&=&\! \!\!\! \fr{1}{r^2}\left(6b_2c_2^2-3b_1c_1^2-2b_3c_3^2\right)-
\fr{1}{2}\lm_1b_2\nonumber\\
&-&\!\!\!
\fr{4\lm_2}{5}\,b_2\left[b_0^2+\fr{3}{2}\left(b_1^2+b_2^2\right)+b_3^2+
b_0\left(b_2+\fr{3b_1}{2}+\fr{b_3}{2}\right)+b_3\left(b_1+\fr{3b_2}{2}\right)+
2b_1b_2\right]\nonumber\\
&-&\!\!\!\fr{\lm_3}{25}\,b_2\left[3b_0^2+12b_1^2+7b_2^2+3b_3^2+6b_0
\left(2b_1+\fr{b_2}{2}-b_3\right)-3b_3\left(4b_1+b_2\right)+6b_1 b_2
\right],\nonumber\acc
\fr{\left(r^2\,AB\,b_3'\right)'}{2Ar^2}
\!\!\!\!&=&\! \!\!\!\fr{1}{r^2}\left(4b_3c_3^2-3b_2c_2^2\right)-\fr{1}{2}
\lm_1b_3
\nonumber\\
&-&\!\!\!
\fr{4\lm_2}{5}\,b_3\left[b_0^2+\fr{3}{2}\left(b_1^2+b_2^2\right)+b_3^2+
b_3\left(b_1+\fr{3b_2}{2}+\fr{b_0}{2}\right)+b_0\left(b_2+\fr{3b_1}{2}\right)+
2b_1b_2\right]\nonumber\\
&-&\!\!\!\fr{\lm_3}{25}\,b_3\left[13b_3^2+27b_2^2+12b_1^2+3b_0^2+18b_3
\left(b_1+\fr{3b_2}{2}+\fr{b_0}{2}\right)+18b_2\left(b_0+2b_1\right)+12b_1 b_0
\right].\nonumber\\
\eea
Finally, the Einstein equations (\ref{ff}) take the form:
\bea
\fr{2}{r^2}\,m'\!\!\!&=&\!\!\!8\pi G\, T_0^0,\\
\fr{1}{r}\fr{A'}{A}\!\!\!&=&\!\!\!8\pi G\Bigg\{\fr{4}{r^2}
\left[c_0'^2+\fr{3}{2}
\left(c_1'^2+c_2'^2\right)+c_3'^2\right]\nonumber\\
&&+\fr{4}{5}\left[b_0'^2+\fr{3}{2}(b_1'^2+b_2'^2)+b_3'^2+
b_0'\left(b_2'+\fr{3b_1'}{2}+\fr{b_3'}{2}\right)+b_3'\left(b_1'+\fr{3}{2}b_2'
\right)
+2b_1'b_2'\right]\Bigg\}\nonumber\\
\eea
where $m(r)$ and $T_0^0$ are given by (\ref{metric}) and (\ref{ene}-\ref{pot}),
respectively.

The above system of equations has to be solved with specific boundary
conditions which ensure the regularity of the solutions and the
finiteness of the ADM mass  defined as: $M_{ADM} = m(\infty) / \alpha^2$.
The Einstein equations imposes the following  boundary
conditions for the  metric  functions:
$m(0)=0$ and $A(\infty)=1$.
The latter condition fixes the invariance of the
equations under the arbitrary scale
$A(r) \rightarrow k\, A(r)$ (for $k$  constant)
and implies that space-time is asymptotically flat.
On the other hand, the regularity of the matter fields
at the origin requires $c_j(0) = 1$ and  $b_j(0) = 0$
while  the finiteness of the ADM mass implies
$b_j(\infty) c_j(\infty)=0$.
 However,
the specific choice of the boundary conditions on $b_j(\infty)$ and
$c_j(\infty)$ is determined by  the type of solution
(e.g. maximal or minimal symmetry breaking) we are interested in.

It is  worth mentioning that, in absence of potential,
the ``length" of the Higgs fields is not fixed since
when $\Phi \rightarrow \lambda \Phi$  and  $r \rightarrow r / \lambda$
 the  ADM mass  scales according to
\be
      M_{ADM}(\lambda \Phi) = \lambda \,M_{ADM}(\Phi).
\ee
This is true also in the flat limit (i.e. for $\alpha = 0$) where
the ADM mass is interpreted as the classical energy of the
 solution.

\section{Numerical Results}
\subsection{Maximal Symmetry Breaking Solutions}

First, we  discuss  solutions with maximal   $SU(5)$ symmetry
breaking that is when all the eigenvalues of $\Phi_0$  (or any
permutation) are  different. 
Since there are many possibilities (in fact, 120
possible permutations exists) we limit ourselves to  few generic
cases.

The simplest case corresponds to the self-dual (SD) solution
(i.e. solution of the Bogomolny equations) where
$\Phi_0 = {\rm diag}(2,1,0,-1,-2)$ implying that the
monopole masses are equal to unity since
$b_0(\infty) = b_1(\infty)= b_2(\infty) = b_3(\infty)=1$ while
the gauge functions $c_j(r)$ vanish  asymptotically.
Then, the  magnetic charge  is $
G_0 = {\rm diag}(4,2,0,-2,-4)$ i.e. $(n_1,n_2,n_3,n_4)=(4,6,6,4)$
 and the corresponding  mass is  equal to $
   M_{max}^{SD} = \sum_{j=1}^{4} b_j n_j = 20$.

Another choice would be:
$b_0(\infty) =b_3(\infty)=3$,  $b_1(\infty)=b_2(\infty) =-2$,
which corresponds to  the non self-dual solution (NSD) where
$G_0 = {\rm diag}(2,-4,0,4,-2)$ 
implying that  $(n_1,n_2,n_3,n_4)=(2,-2,-2,2)$.
As expected, the solution cannot be constructed analytically since the
corresponding equations are not integrable. However,  it can be obtained
numerically and  its mass
is evaluated to be  equal to  $M_{max}^{NSD} = 27$.

\begin{figure}[c]
\centering
\epsfysize=18cm
\mbox{\epsffile{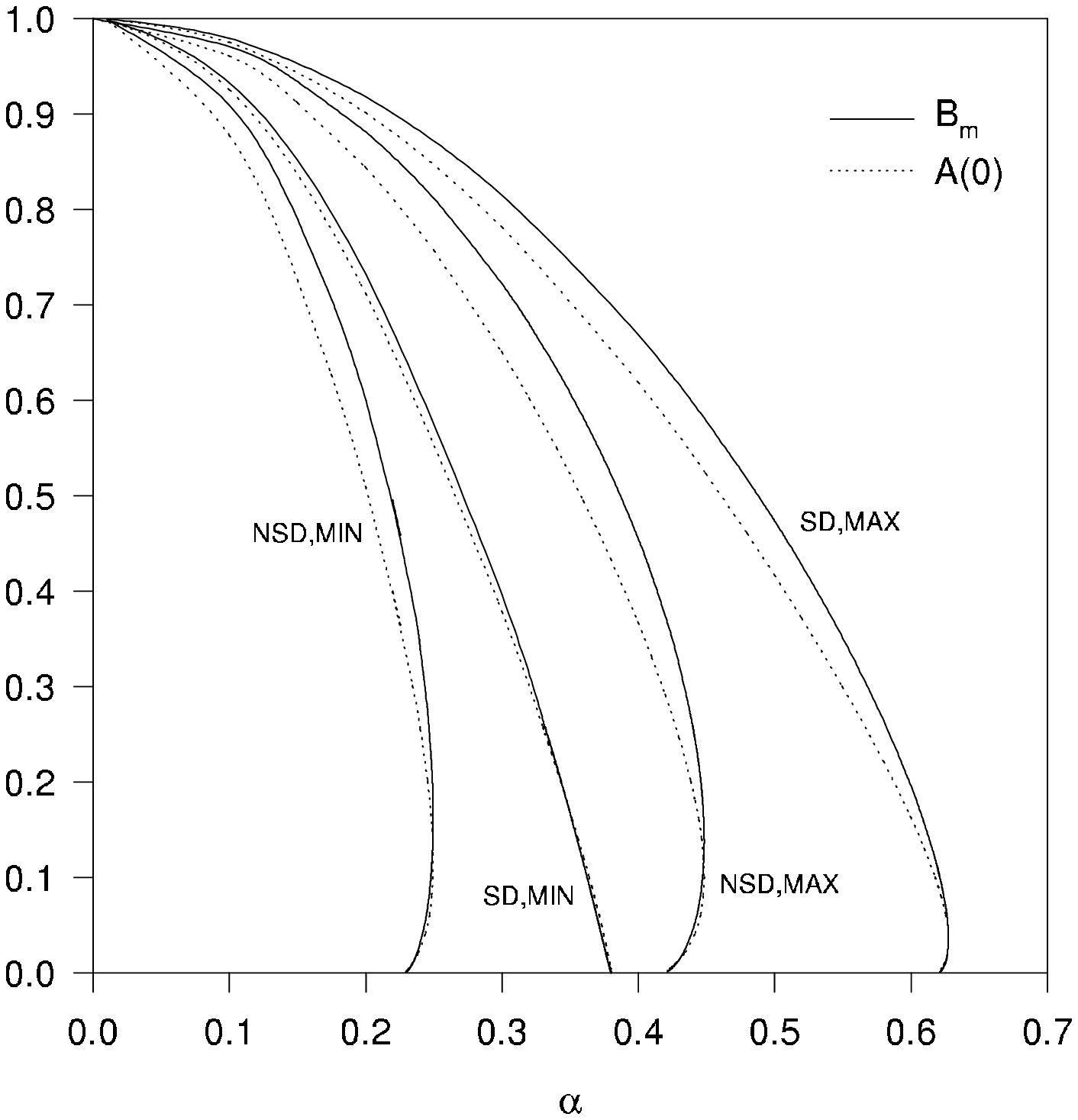}}
\caption{\label{Fig.1} The 
metric functions $A(0)$ and $B_m=\mbox{min}[B(r)]$ in terms of $\al$.}
\centering
\end{figure}

\begin{figure}[c]
\centering
\epsfysize=18cm
\mbox{\epsffile{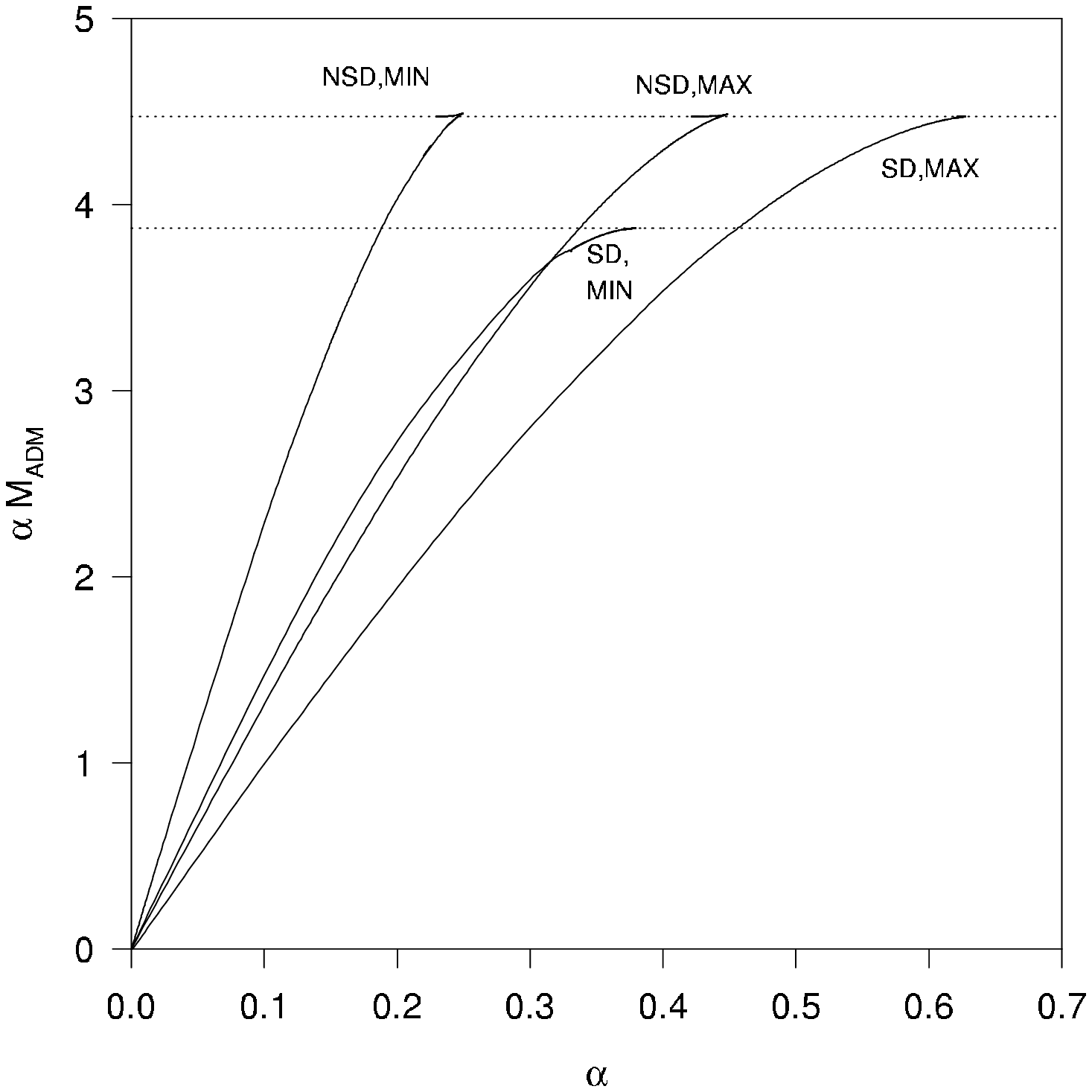}}
\caption{\label{Fig.2} The ADM mass in terms  of the coupling constant $\al$. }
\centering
\end{figure}

When these solutions are coupled to gravity
(i.e. $\al \neq 0$),
numerical simulations indicate that they  deformed to
 ``gravitating" $SU(5)$ monopoles while their presence
  progressively deforms space-time. For
instance, the function $B(r)$ develops a minimum $B=B_m$ at some
intermediate value of  $r=r_h$,
 as  $\al$ increases.
At the same time,  $A(r)$   takes its minimum
value at the origin while  at infinity tends  to the value $A(\infty)=1$.
The metric functions $A(0$) and $B_m$ decrease  as  $\al$ increases
for the self-dual (line SD, MAX) and non self-dual
(line NSD, MAX) solutions, as indicated in Figure \ref{Fig.1}.
Similarly, in Figure \ref{Fig.2} the $\al$ dependence of the 
product $\al M_{ADM}$  is plotted (using the same conventions for
 the various lines) and shows that the  ADM mass and the product 
$\al M_{ADM}$  decreases and increases (respectively) as $\al$ increases.
It should be stressed out that, due to the peculiar normalisation
of the Higgs field, the energies are not directly comparable;
therefore, the ratio
$\al M_{ADM}/ \vert \Phi_0 \vert$ should be considered instead.
Both branches stop at some maximal value
of $\al$ ; i.e. the self-dual solution can be
deformed by gravity up to $\alpha_m \approx 0.63$; while
the non self-dual one
exists up to $\alpha_m \approx 0.45$.
As in  the $SU(3)$ case \cite{BHIZ}, 
the region of $\al$  in order  gravitating monopoles to exist
decreases as  the mass of the flat solution increases.
However this is not the end of the story.
The main branches  of gravitating solutions (i.e. the non-gravtitating ones)
are completed by secondary
branches  which exist on a rather small interval
of $\al$, as seen in the  $SU(2)$
\cite{bfm} and  $SU(3)$ \cite{BHIZ} models.
Indeed, the  secondary branch exists in the following intervals:
 \begin{eqnarray}
&\mbox{SD}, \mbox{MAX}: \ \
&\alpha_{cr} \approx 0.621, \ \  \alpha_m \approx 0.627
\nonumber \\
&\mbox{NSD}, \mbox{MAX}: \ \
&\alpha_{cr} \approx 0.424, \ \  \alpha_m \approx 0.448
 \end{eqnarray}
and  as $\alpha \rightarrow \alpha_{cr}$  the
minimum of $B(r)$ tends to  zero which means (in this limit) the
solution develops a horizon (shown in  Figure \ref{Fig.1}).
 In fact, $B_m\ra 0$  faster than $A(0)$ in terms of the critical value of
$\al$ which means that the $SU(5)$ gravitating monopole
bifurcates into an extremal Reissner-Nordstrom  black hole.
This configuration corresponds to solutions of the abelian
Einstein-Maxwell equations and can be embedded into the non-abelian ones.
Its mass  is equal to
\be
      m_{RN} = m_{\infty,RN} - \frac{ \alpha^2 Q^2}{2r}
\ee
where $Q$ is the charge of the black hole
and  can be read off from the energy-momentum tensor, i.e.
\be
\frac{Q^2}{2}\!\!=\!\!
\left[8(c_0^4+c_3^4)+18\left(c_1^4+c_2^4\right)-4(c_0^2+c_3^2)-6
\left(c_1^2+c_2^2\right)
-12\left(c_1^2c_0^2+c_2^2c_3^2\right)-18c_1^2c_2^2+10\right]
\bigg|_{r=\infty}.
\ee
Both our  solutions  have charge equal to $Q= \sqrt{20}$ in consistence
with the  numerical simulations (see Figure \ref{Fig.2}).

\subsection{Minimal Symmetry Breaking Solutions}

The $SU(5)$ symmetry can be broken into many  minimal breaking
patterns producing solutions with  non-abelian stability group. In
what follows, we present two types of such solutions
 which are invariant under the $SU(3) \times SU(2)/U(1)$ group.

\begin{figure}
\centering
\epsfysize=18cm
\mbox{\epsffile{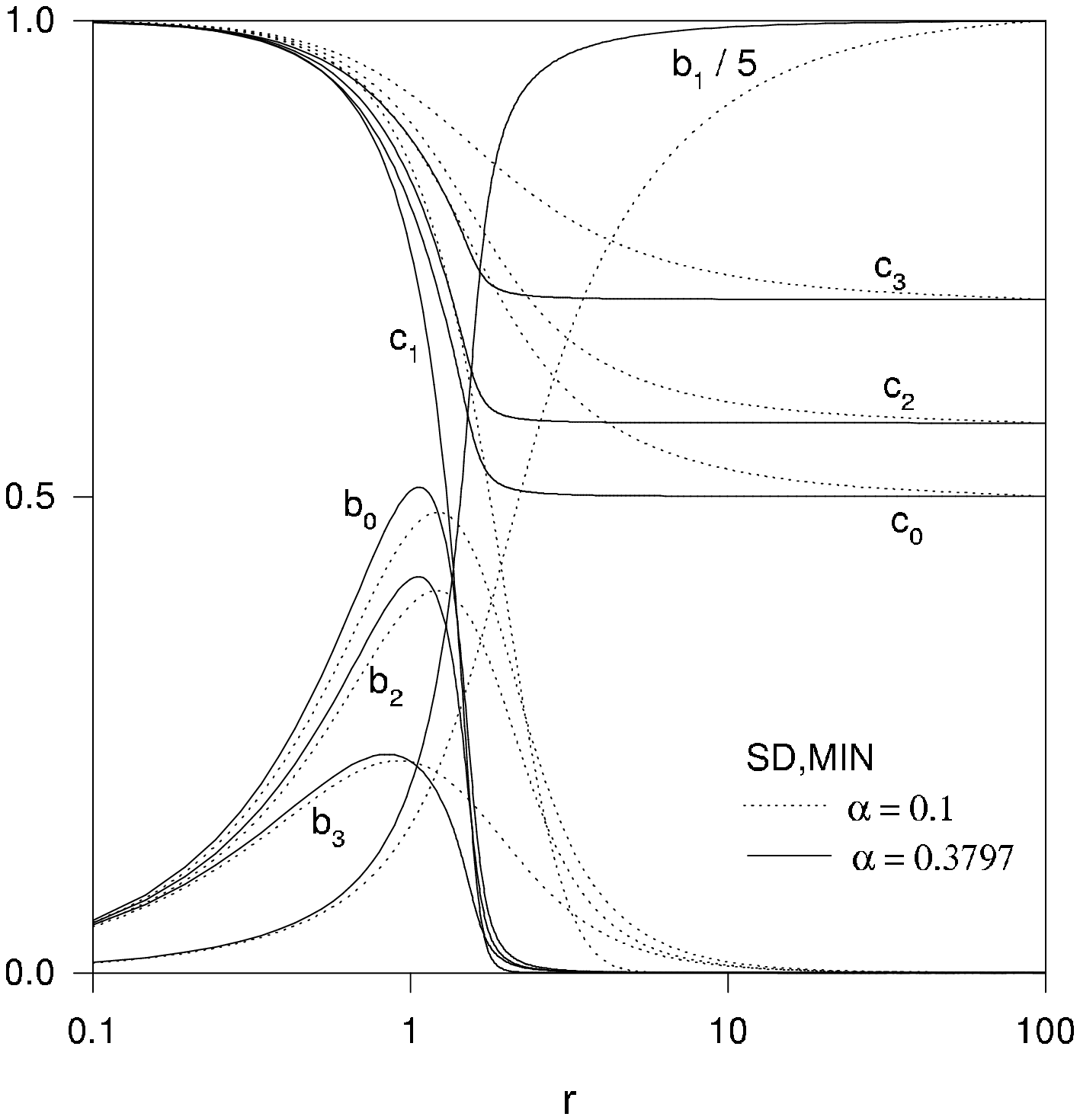}}
\caption{\label{Fig.3} The matter profile functions
of the self-dual minimal symmetry breaking
 solution for two different values of $\al$.}
\centering
\end{figure}

First, we investigate the self-dual solution where
$ \Phi_0 = {\rm diag} (3,3,-2,-2,-2)$
i.e. for  $ b_0(\infty)=b_2(\infty)=b_3(\infty)=0$ and $b_1(\infty)=5$ 
while the $c_j$
fields satisfy the following asymptotic values
$c_0 = \frac{1}{2}$,
$c_1 = 0$,
$c_2 = \frac{1}{\sqrt{3}}$ and $c_3 = \frac{1}{\sqrt{2}}$.
The corresponding solution has energy  equal to $E = 30$ with  mass
(or classical energy in the flat limit) lower
compare to  all types of solutions we investigated -
when normalized appropriately. When it is coupled with gravity
 it exists up to
$\alpha_m \approx 0.38$. Contrary to the other cases, 
 this solution (on the main branch) bifurcates into a Reissner-Nordstrom
solution with charge $Q = \sqrt{15}$ as  confirmed by  our
numerical analysis which  does not indicate the
existence  of a second branch. Figure \ref{Fig.3} illustrates the
way the  matter functions approach  their (constant) values
outside the horizon (at $r=\alpha_{cr}$) of the approached
Reissner-Nordstrom solution when $\al = 0.1$    (i.e. close to the
flat limit) and    $\al = 0.3797$ (i.e. close to the critical limit).

Finally,  another non self-dual solution with  the same unbroken group can
be constructed   when $\Phi_0$ is of the form 
$\Phi_0 = {\rm diag} (2,-3,2,-3,2)$ 
i.e. for   $b_0(\infty)=b_2(\infty)=1$, $b_1(\infty)=b_3(\infty)=-1$ 
 and  $c_j(\infty)=0$.
The corresponding mass of the configuration is equal to $M=48$.

Once more, the aforementioned solutions can be considered in the
presence of gravitating fields and our numerical routines indicate
that their gravitating analogues exist up to a maximal value of
the coupling constant equal to
  $\alpha_m \approx 0.249$.
In addition,  a secondary branch exists  which  terminates at
$\al \approx 0.229$ into an extremal Reissner-Nordstrom black
hole of charge $Q = \sqrt{20}$.
The corresponding results are presented in Figure \ref{Fig.1} and Figure
\ref{Fig.2} (line SD, MIN) and (line NSD, MIN).

\section{Conclusions}

In this paper, four types of $SU(5)$ monopoles have been
constructed which  can be deformed by gravity  forming branches of
solutions labelled by the gravitational coupling constant
$\alpha$. Numerical investigation of the  $SU(5)$
Einstein-Yang-Mills-Higgs equations reveals  that for each branch,
 a second one exists  on the  interval $\alpha \in
[\alpha_{cr}, \alpha_m$] (depending on its type) 
in consistence with the
results obtained in  smaller gauge groups like $ SU(2)$ and
$SU(3)$. In fact, the solution on the second branch has a higher
mass than the one with the same $\alpha$ on the first (or main)
branch; while, in the limit $\alpha \rightarrow \alpha_{cr}$, the
minimum of the function $B(r)$ becomes deeper and deeper and
approaches zero at some intermediate value $r_h$. Accordingly, the
regular solution does not exist for $\alpha = \alpha_{cr}$ and the
metric fields approach that of  an extremal Reissner-Nordstrom
black hole on the interval $[r_h,\infty]$
while  all matter fields tend to their
asymptotic values.\\

{\bf Acknowledgements}
We gratefully acknowledge Betti Hartmann for numerous
interesting discussions which turn out to be
at the basis of the present work.
Y.B. acknowledges the Belgian F.N.R.S. for financial support.

\end{document}